\documentclass[12pt]{article}
\usepackage{graphics}

\title{$CP$- and $T$-Violation in the Decay\\ $K_L \rightarrow \pi^+ \pi^- e^+ e^-$
and Related Processes\thanks{Talk given at the KAON 99 Conference, Chicago,
June 21-26, 1999}}

\author{L. M. Sehgal \\ \it{Institute of Theoretical Physics, RWTH Aachen,} \\
\it{D-52056 Aachen, Germany}}

\date{}
\begin{document}

\maketitle

\begin{abstract}
I review the theoretical basis of the prediction that the decay
$K_L \rightarrow \pi^+ \pi^- e^+ e^-$ should show a large $CP$- and $T$-violation,
a prediction now confirmed by the KTeV experiment. The genesis of the effect
lies in a large violation of $CP$- and $T$-invariance in the decay
$K_L \rightarrow \pi^+ \pi^- \gamma$, which is encrypted in the polarization
state of the photon. The decay $K_L \rightarrow \pi^+ \pi^- e^+ e^-$ serves
as an analyser of
the photon polarization. The asymmetry in the distribution of the angle $\phi$
between the $\pi^+ \pi^-$ and $e^+ e^-$ planes is a direct measure of the $CP$-odd,
$T$-odd component of the photon's Stokes vector. A complete study of the angular
distribution can reveal further $CP$-violating features, which probe
the non-radiative (charge-radius and short-distance) components of the
$K_L \rightarrow \pi^+ \pi^- e^+ e^-$ amplitude.
\end{abstract}

Eight years ago, there appeared a report~\cite{E731coll} by the E-731 experiment
concerning the branching ratio and photon energy spectrum of the 
decays $K_{L,S} \rightarrow \pi^+ \pi^- \gamma$. It was found that while the
$K_S$ decay could be well-reproduced by inner bremsstrahlung from an underlying
process $K_S \rightarrow \pi^+ \pi^-$, the $K_L$ decay contained a mixture of
a bremsstrahlung component ($I\!B$) and a direct emission component ($D\!E$), the
relative strength being $D\!E/(D\!E+I\!B) = 0.68$ for photons above $20\, MeV$.
The simplest matrix element consistent with these features is
\begin{eqnarray}
\mathcal{M} (K_S \rightarrow \pi^+ \pi^- \gamma) & = & e f_S
\left[\frac{\epsilon \cdot p_+}{k \cdot p_+} -
\frac{\epsilon \cdot p_-}{k \cdot p_-} \right] 
\label{matelem}\\
\mathcal{M} (K_L \rightarrow \pi^+ \pi^- \gamma) & = & e f_L
\left[\frac{\epsilon \cdot p_+}{k \cdot p_+} -
\frac{\epsilon \cdot p_-}{k \cdot p_-} \right] + e \frac{f_{DE}}{{M_K}^4}
\epsilon_{\mu\nu\rho\sigma} \epsilon^{\mu}k^{\nu}{p_+}^{\rho}{p_-}^{\sigma}
\nonumber
\end{eqnarray}
where
\begin{eqnarray}
f_L \equiv |f_S| g_{Br},\: g_{Br} = \eta_{+-} e^{i \delta_0(s={M_K}^2)},
\nonumber\\
f_{DE} \equiv |f_S| g_{M1},\: g_{M1} = i(0.76)e^{i \delta_1(s)}.
\label{whatmeansf}
\end{eqnarray}
Here the direct emission has been represented by a $CP$-conserving magnetic
dipole coupling $g_{M1}$, whose magnitude $|g_{M1}| = 0.76$ is fixed by the
empirical ratio $DE/IB$. The phase factors appearing in $g_{Br}$ and $g_{M1}$
are dictated by the Low theorem for bremsstrahlung,
and the Watson theorem for final state interactions. The factor $i$ in
$g_{M1}$ is a consequence of $CPT$ invariance~\cite{Sehgal:Costa:Lee}.
The matrix element for $K_L \rightarrow \pi^+ \pi^- \gamma$ contains 
simultaneously electric multipoles associated with
bremsstrahlung ($E1,\,E3,\,E5$ ...), which have $CP=+1$, and a magnetic $M1$
multipole with $CP = -1$. It follows that interference of the electric and
magnetic emissions should give rise to $CP$-violation.

To understand the nature of this interference, we write the
$K_L \rightarrow \pi^+ \pi^- \gamma$ amplitude more generally as
\begin{eqnarray}
\mathcal{M}(K_L \rightarrow \pi^+ \pi^- \gamma) & = &
\frac{1}{{M_K}^3} \left\{ E(\omega,\cos \theta) \left[\epsilon \cdot p_+ \, k
\cdot p_- - \epsilon \cdot p_- \, k \cdot p_+ \right] \right. \nonumber\\
& & \left. \mbox{} + M(\omega, \cos \theta)
\epsilon_{\mu\nu\rho\sigma}\epsilon^{\mu}k^{\nu}{p_+}^{\rho}{p_-}^{\sigma}
\right\}
\end{eqnarray}
where $\omega$ is the photon energy in the $K_L$ rest frame, and $\theta$
is the angle between $\pi^+$ and $\gamma$ in the $\pi^+ \pi^-$ rest frame.
In the model represented by Eqs. (\ref{matelem}) and (\ref{whatmeansf}),
the electric and magnetic amplitudes are
\begin{eqnarray}
E & = & \left( \frac{2M_K}{\omega} \right)^2 \frac{g_{Br}}{1-\beta^2 \cos^2 \theta}
\nonumber\\
M & = & g_{M1}
\label{defEandM}
\end{eqnarray}
where $\beta = (1- 4 {m_{\pi}}^2/s)^{1/2}$, $\sqrt{s}$ being the $\pi^+\pi^-$ invariant
mass. The Dalitz plot density, summed over photon polarizations is
\begin{equation}
\frac{d \Gamma}{d \omega \, d\! \cos \theta} = \frac{1}{512 \pi^3}
\left( \frac{\omega}{M_K} \right)^3 \beta^3 \left( 1- \frac{2 \omega}{M_K}
\right) \sin^2 \theta \left[ |E|^2 + |M|^2 \right]
\label{dGammaint}
\end{equation}
Clearly, there is no interference between the electric and magnetic multipoles
if the photon polarization is unobserved. Therefore, any $CP$-violation
involving the interference of $g_{Br}$ and $g_{M1}$ is hidden in the
polarization state of the photon. 

The photon polarization can be defined in terms of the density matrix
\begin{equation}
\rho = \left( \begin{array}{cc} |E|^2 & E^*M \\ EM^* & |M|^2 \end{array} 
\right) = \frac{1}{2} \left( |E|^2+|M|^2 \right) \left[ 1 \!\!\!\!\:\: \mbox{l}
 + \vec{S} \cdot \vec{\tau}  \right]
\end{equation}
where $\vec{\tau} = (\tau_1, \, \tau_2, \, \tau_3)$ denotes the Pauli matrices,
and $\vec{S}$ is the Stokes vector of the photon with components
\begin{eqnarray}
S_1 & = & 2 Re \left( E^*M \right) / \left( |E|^2 + |M|^2 \right) \nonumber\\
S_2 & = & 2 Im \left( E^*M \right) / \left( |E|^2 + |M|^2 \right) \\
S_3 & = & \left(|E|^2 - |M|^2 \right) / \left( |E|^2 + |M|^2 \right). \nonumber
\end{eqnarray}
The component $S_3$ measures the relative strength of the electric and magnetic
radiation at a given point in the Dalitz plot. The effects of $CP$-violation
reside in the components $S_1$ and $S_2$, which are proportional to
$Re \, ({g_{Br}}^* g_{M1})$ and $Im \, ({g_{Br}}^* g_{M1})$, respectively. Of
these $S_1$ is $CP$-odd, $T$-odd, while $S_2$ is $CP$-odd, $T$-even. Physically,
$S_2$ is the net circular polarization of the photon: such a polarization
is known to be possible in decays like $K_L \rightarrow \pi^+ \pi^- \gamma$ or
$K_{L,S} \rightarrow \gamma \gamma$ whenever there is $CP$-violation accompanied
by unitarity phases~\cite{Sehgal}. To understand the significance of $S_1$,
we examine the dependence of the $K_L \rightarrow \pi^+ \pi^- \gamma$ decay
on the angle $\phi$ between the polarization vector $\vec{\epsilon}$ and
the unit vector $\vec{n}_{\pi}$ normal to the $\pi^+ \pi^-$ plane (we
choose coordinates such that $\vec{k} = (0,\, 0,\, k)$, $\vec{n}_{\pi} =
(1,\, 0, \, 0)$, $\vec{p}_+ = (0,\, p\sin \theta, \, p\cos \theta)$ and
$\vec{\epsilon} = (\cos \phi, \, \sin \phi, \, 0)$):
\begin{equation}
\frac{d \Gamma}{d \omega \, d\! \cos \theta \, d \phi} \sim
\left| E \sin \phi - M \cos \phi \right|^2 \sim
1 - \left[ S_3 \cos 2\phi + S_1 \sin 2\phi \right]
\end{equation}
Thus the $CP$-odd, $T$-odd Stokes parameter $S_1$ appears as a coefficient
of the term $\sin 2\phi$. The essential idea of Refs.~\cite{Sehgal:Wanninger,Heiliger:Sehgal}
is to use in place of $\vec{\epsilon}$,
the vector $\vec{n}_l$ normal to the plane of the Dalitz pair in the reaction
$K_L \rightarrow \pi^+ \pi^- \gamma^* \rightarrow \pi^+ \pi^- e^+ e^-$. This
motivates the study of the distribution $d\Gamma/d\phi$ in the decay
$K_L \rightarrow \pi^+ \pi^- e^+ e^-$, where $\phi$ is the angle between
the $\pi^+ \pi^-$ and $e^+ e^-$ planes.

\begin{figure}[ht]
\makebox[15.8pc]{
\resizebox{15.8pc}{14.4pc}
{\includegraphics*[2.5cm,2.5cm][27.7cm,18.5cm]{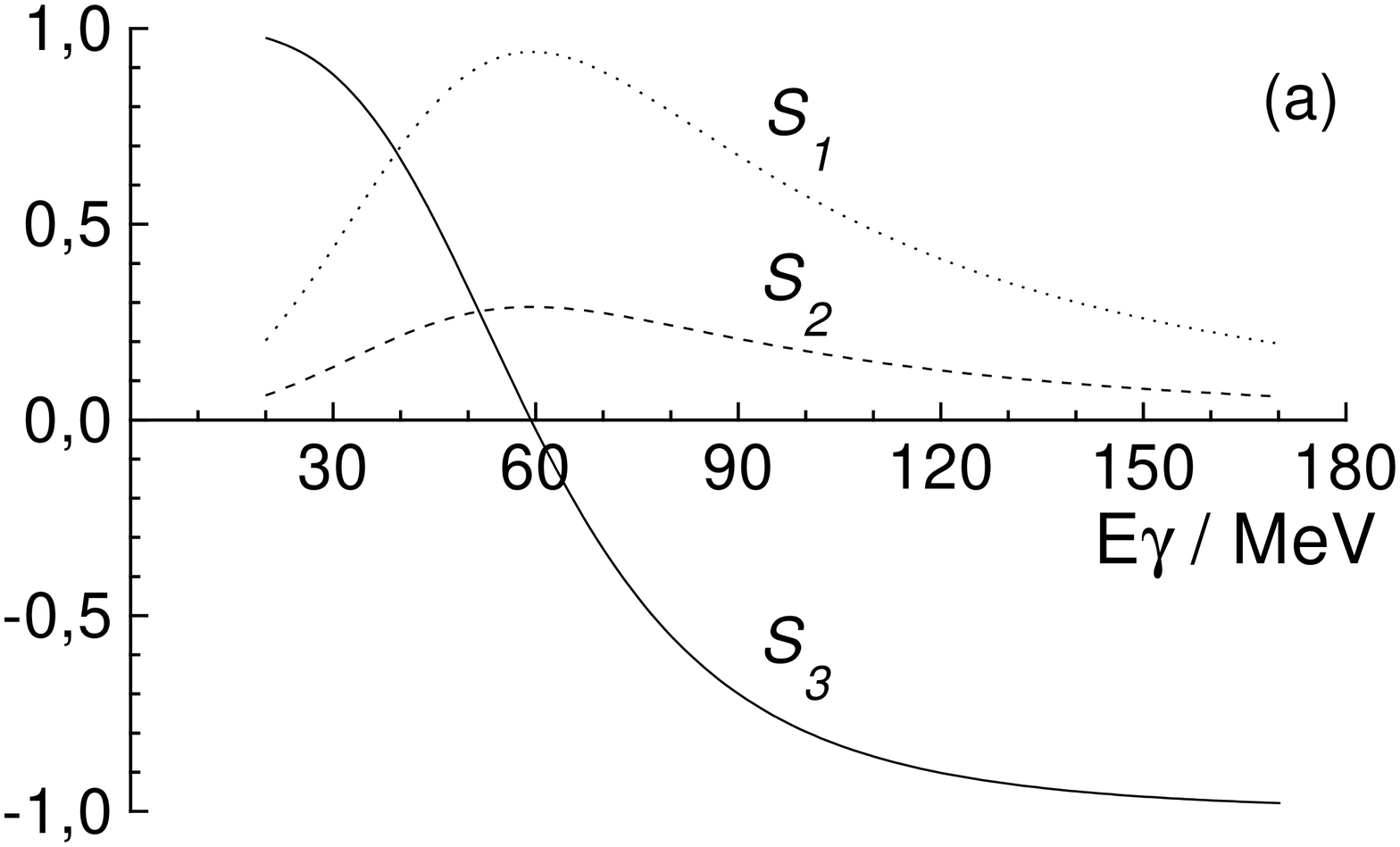}}
}
\makebox[15.8pc]{
\resizebox{15.8pc}{14.4pc}
{\includegraphics*[2.5cm,2.5cm][27.7cm,18.5cm]{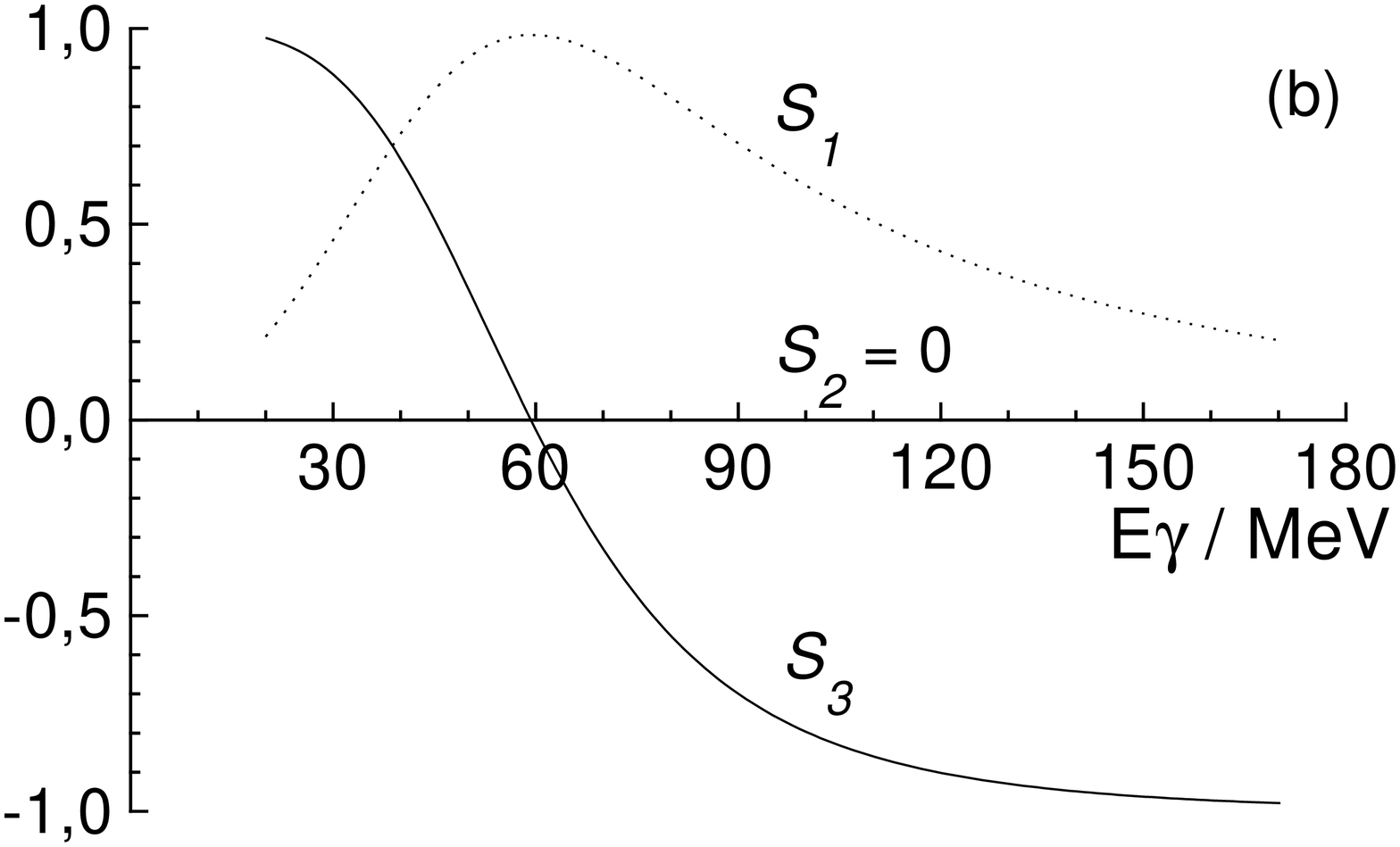}}
}
\makebox[15.8pc]{
\resizebox{15.8pc}{14.4pc}
{\includegraphics*[2.5cm,2.5cm][27.7cm,18.5cm]{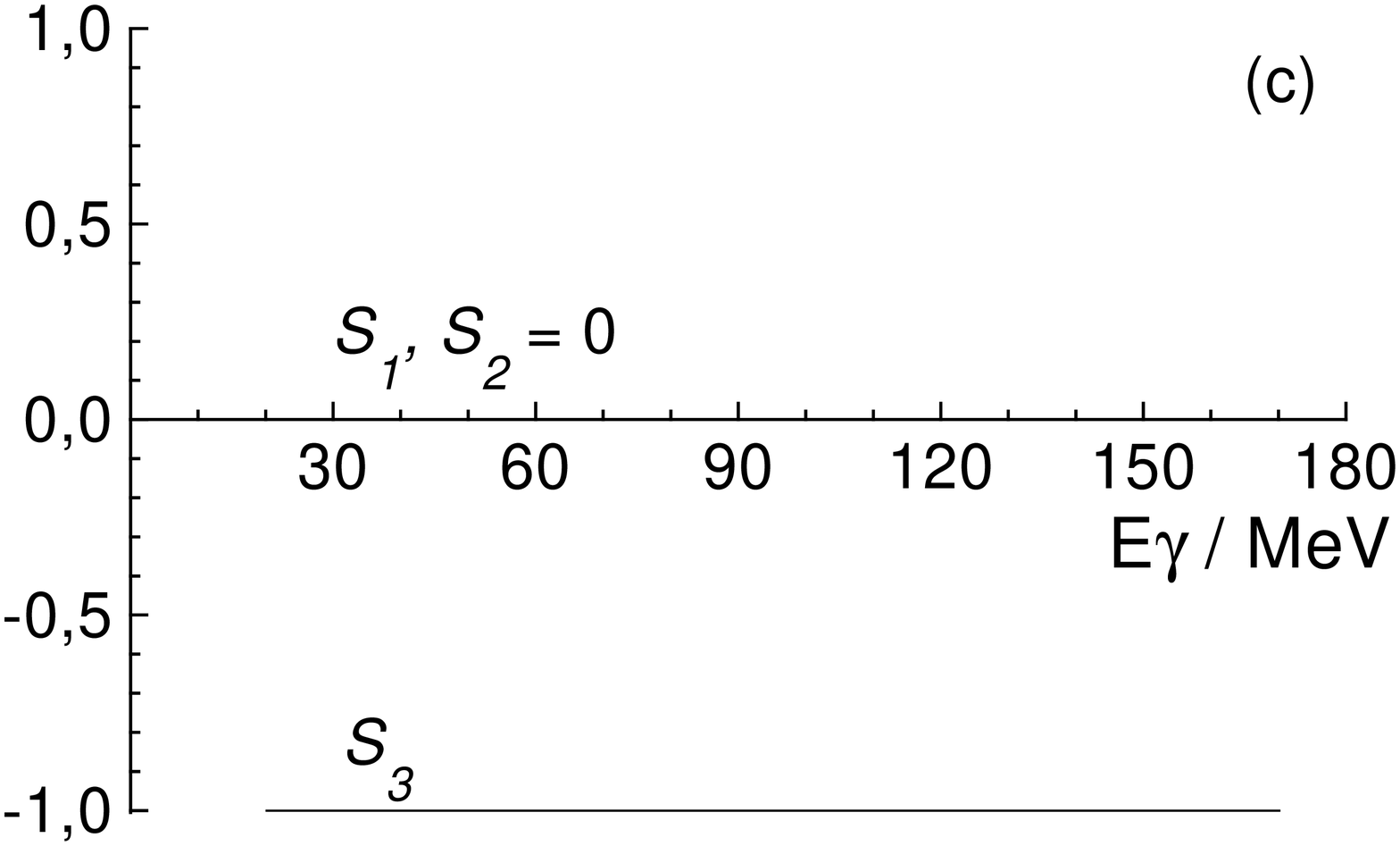}}
}
\makebox[15.8pc]{
\resizebox{15.8pc}{14.4pc}
{\includegraphics*[2.5cm,2.5cm][27.7cm,18.5cm]{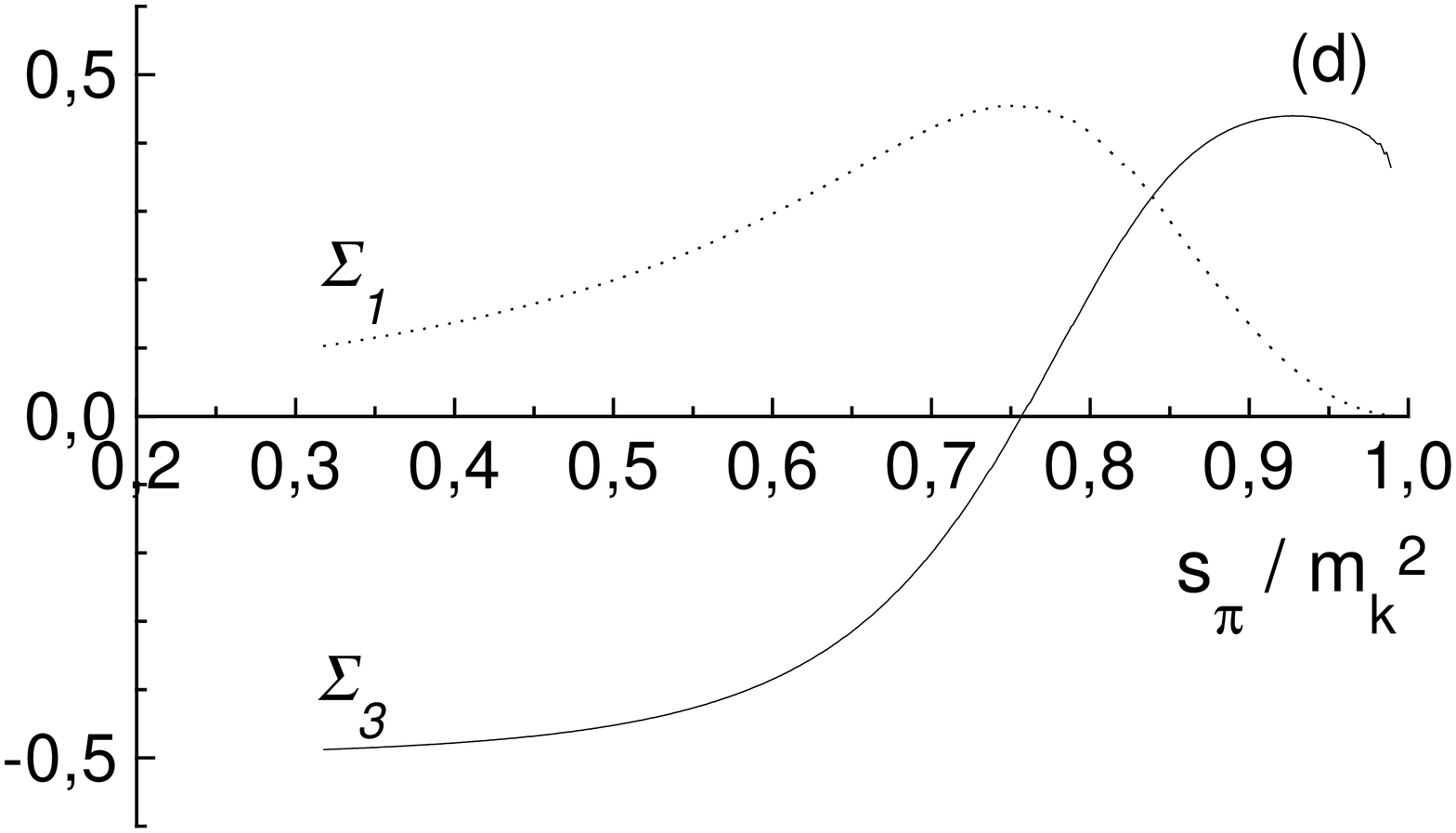}}
}
\caption{(a) Stokes parameters of photon in 
$K_L \rightarrow \pi^+ \pi^- \gamma$; (b) Hermitian limit 
$\delta_0 = \delta_1 = 0$, $ar\!g \, \epsilon = \pi / 2$;
(c) $CP$-invariant limit $\epsilon \rightarrow 0$;
(d) ``Stokes parameters'' for $K_L \rightarrow \pi^+ \pi^- e^+ e^-$.
\label{SivsE}
}
\end{figure}

To obtain a quantitative idea of the magnitude of $CP$- violation in
$K_L \rightarrow \pi^+ \pi^- \gamma$, we show in Fig.~\ref{SivsE}a the three
components of the Stokes  vector as a function of the photon
energy~\cite{Sehgal:vanLeusen}. These are calculated from the amplitudes
(\ref{defEandM}) using weighted averages of $|E|^2$, $|M|^2$, $E^*M$
and $EM^*$ over $\cos \theta$. The values of $S_1$ and $S_2$
are remarkably large, considering that the only source of $CP$-violation is
the $\epsilon$-impurity in the $K_L$ wave-function ($\epsilon = \eta_{+-}$).
Clearly the $1/\omega^2$ factor in $E$ enhances it to a level
that makes it comparable to the $CP$-conserving amplitude $M$. This is
evident from the behaviour of the parameter $S_3$, which swings from a
dominant electric behaviour at low $E_{\gamma}$ ($S_3 \approx 1$) to a
dominant magnetic behaviour at large $E_{\gamma}$ ($S_3 \approx -1$),
with a zero in the region $E_{\gamma} \approx 60 \, MeV$. To highlight the
difference between the $T$-odd parameter $S_1$ and the $T$-even parameter
$S_2$, we show in Fig.~\ref{SivsE}b the behaviour of the Stokes parameters
in the ``hermitian limit'': this is the limit in which the $T$-matrix or
effective Hamiltonian governing the decay $K_L \rightarrow \pi^+ \pi^- \gamma$
is taken to be hermitian, all unitarity phases related to real intermediate
states being dropped. This limit is realized by taking
$\delta_0 , \, \delta_1 \rightarrow 0$, and $ar\!g \, \epsilon \rightarrow \pi /2$.
The last of these follows from the fact that $\epsilon$ may be written as
\begin{equation}
\epsilon = \frac{\Gamma_{12}-\Gamma_{21} + i \left( M_{12}-M_{21} \right)}
{\gamma_S - \gamma_L - 2 i \left( m_L - m_S \right)}
\end{equation}
where $H_{eff} = M - i\Gamma$ is the mass matrix of the $K^0$-$\overline{K}^0$
system. The hermitian limit obtains when $\Gamma_{12} = \Gamma_{21} = \gamma_S
= \gamma_L = 0$. As seen from Fig.~\ref{SivsE}b, $S_2$ vanishes in this
limit, but $S_1$ survives, as befits a $CP$-odd, $T$-odd parameter.
Fig.~\ref{SivsE}c shows what happens in the $CP$-invariant limit 
$\epsilon \rightarrow 0$.
It is clear that we are dealing here with a dramatic 
situation in which a $CP$-impurity of a few parts in a thousand
in the $K_L$ wave-function gives rise to a huge $CP$-odd, $T$-odd effect
in the photon polarization.

We can now examine how these large $CP$-violating effects are transported
to the decay $K_L \rightarrow \pi^+ \pi^- e^+ e^-$. The matrix element for
$K_L \rightarrow \pi^+ \pi^- e^+ e^-$ can be written as~\cite{Sehgal:Wanninger,Heiliger:Sehgal}
\begin{equation}
\mathcal{M} (K_L \rightarrow \pi^+ \pi^- e^+ e^-) = \mathcal{M}_{br} +
\mathcal{M}_{mag} + \mathcal{M}_{CR} + \mathcal{M}_{SD}.
\end{equation}
Here $\mathcal{M}_{br}$ and $\mathcal{M}_{mag}$ are the conversion amplitudes
associated with the bremsstrahlung and $M1$ parts of the
$K_L \rightarrow \pi^+ \pi^- \gamma$ amplitude. In addition, we have
introduced an amplitude $\mathcal{M}_{CR}$ denoting $\pi^+ \pi^-$ production
in a $J = 0$ state (not possible in a real radiative decay), as well as an
amplitude $\mathcal{M}_{SD}$ associated with the short-distance interaction
$s \rightarrow d\, e^+ e^-$. The last
of these turns out to be numerically negligible because of the smallness
of the $C\!K\!M$ factor $V_{ts} {V_{td}}^*$. The $s$-wave amplitude
$\mathcal{M}_{CR}$, if approximated by the $K^0$ charge radius diagram,
makes a small ($\sim 1 \%$) contribution to the decay rate. Thus the
dominant features of the decay are due to the conversion amplitude
$\mathcal{M}_{br} + \mathcal{M}_{mag}$.

Within such a model, one can calculate the differential decay rate in the
form~\cite{Heiliger:Sehgal}
\begin{equation}
d \Gamma = I(s_{\pi}, \, s_l, \, \cos \theta_l, \, \cos \theta_{\pi}, \, \phi)
\, ds_{\pi} \, ds_l \, d\! \cos \theta_l \, d\! \cos \theta_{\pi} \, d\phi.
\end{equation}
Here $s_{\pi}$ ($s_l$) is the invariant mass of the pion (lepton) pair,
and $\theta_{\pi}$ ($\theta_l$) is the angle of the $\pi^+$ ($l^+$) in the
$\pi^+ \pi^-$ ($l^+ l^-$) rest frame, relative to the dilepton (dipion)
momentum vector in that frame. The all-important variable $\phi$ is defined
in terms of unit vectors constructed from the pion momenta $\vec{p_{\pm}}$
and lepton momenta $\vec{k_{\pm}}$ in the $K_L$ rest frame:
\begin{eqnarray}
\vec{n}_{\pi} = \left( \vec{p}_+ \times \vec{p}_- \right) /
\left| \vec{p}_+ \times \vec{p}_- \right|,
& \vec{n}_l = \left( \vec{k}_+ \times \vec{k}_- \right) /
\left| \vec{k}_+ \times \vec{k}_- \right|, \nonumber\\
\vec{z} = \left( \vec{p}_+ + \vec{p}_- \right) /
\left| \vec{p}_+ + \vec{p}_- \right|, \nonumber
\end{eqnarray}
\begin{eqnarray}
\sin \phi \: = & \vec{n}_{\pi} \times \vec{n}_l \cdot \vec{z} & (CP = -, T = -), \\
\cos \phi \: = & \vec{n}_l \cdot \vec{n}_{\pi} & (CP = +, T = +). \nonumber
\end{eqnarray}
In Ref.~\cite{Sehgal:Wanninger}, an analytic expression was derived for
the 3-dimensional distribution $d\Gamma/ ds_l \, ds_{\pi} \, d\phi$, which
has been used in the Monte Carlo simulation of this
decay. In Ref.~\cite{Heiliger:Sehgal}, a formalism was presented for
obtaining the fully differential decay function $I(s_{\pi},\, s_l, \, \cos
\theta_l, \, \cos \theta_{\pi}, \phi)$.

The principal results of the theoretical model discussed
in~\cite{Sehgal:Wanninger, Heiliger:Sehgal} are as follows:

1. Branching ratio: This was calculated to be~\cite{Sehgal:Wanninger}
\begin{eqnarray}
B\!R(K_L \rightarrow \pi^+ \pi^- e^+ e^-) & = & (1.3 \times 10^{-7})_{Br} +
(1.8 \times 10^{-7})_{M1} \nonumber \\
 & & \mbox{} + (0.04 \times 10^{-7})_{CR} \approx 3.1 \times 10^{-7},
\end{eqnarray}
which agrees well with the result $(3.32 \pm 0.14 \pm 0.28) \times 10^{-7}$ 
measured in the
KTeV experiment~\cite{KTeVcoll}. (A preliminary branching ratio 
$2.9 \times 10^{-7}$ has been reported by NA48~\cite{NA48coll}).

2. Asymmetry in $\phi$ distribution: The model predicts a distribution of
the form
\begin{equation}
\frac{d\Gamma}{d\phi} \sim 1 - \left( \Sigma_3 \cos 2\phi + \Sigma_1 \sin 2\phi
 \right)
\end{equation}
where the last term is $CP$- and $T$-violating, and produces an asymmetry
\begin{equation}
\mathcal{A} = \frac{\left( \int_{0}^{\pi/2} - \int_{\pi/2}^{\pi} +
\int_{\pi}^{3\pi/2} - \int_{3\pi/2}^{2\pi} \right) \frac{d\Gamma}{d\phi} d\phi}
{\left( \int_{0}^{\pi/2} + \int_{\pi/2}^{\pi} +
\int_{\pi}^{3\pi/2} + \int_{3\pi/2}^{2\pi}\right) \frac{d\Gamma}{d\phi} d\phi}
= - \frac{2}{\pi} \Sigma_1.
\end{equation}
The predicted value~\cite{Sehgal:Wanninger,Heiliger:Sehgal} is
\begin{equation}
|\mathcal{A}| = 15 \% \, \sin (\phi_{+-} + \delta_0({M_K}^2) - \overline{\delta}_1 )
\approx 14 \%
\end{equation}
to be compared with the KTeV result~\cite{KTeVcoll}
\begin{equation}
|\mathcal{A}|_{KTeV} = (13.6 \pm 2.5 \pm 1.2) \%
\end{equation}
The  ``Stokes parameters'' $\Sigma_3$ and $\Sigma_1$ are calculated to be 
$\Sigma_3 = -0.133$, $\Sigma_1 = 0.23$. The $\phi$-distribution
measured by KTeV agrees with this expectation (after acceptance corrections
made in accordance with the model). It should be noted that the sign of
$\Sigma_1$ (and of the asymmetry $\mathcal{A}$) depends on whether the
numerical coefficient in $g_{M1}$ is taken to be $+0.76$ or $-0.76$. The
data happen to support the positive sign chosen in Eq. (\ref{whatmeansf}).

3. Variation of Stokes parameters with $s_{\pi}$: As shown in
Fig.~\ref{SivsE}d, the parameters $\Sigma_1$ and $\Sigma_3$ have a
variation with $s_{\pi}$ that is in close correspondence with the variation
of $S_1$ and $S_3$ shown in Fig.~\ref{SivsE}. (Recall that the photon energy
$E_{\gamma}$ in $K_L \rightarrow \pi^+ \pi^- \gamma$ can be expressed in terms
of $s_{\pi}$: $s_{\pi} = {M_K}^2 - 2M_K E_{\gamma}$.) In particular the zero
of $\Sigma_3$ and the zero of $S_3$ occur at almost the same value of $s_{\pi}$.
This variation with $s_{\pi}$ combined with the low detector acceptance at
large $s_{\pi}$, has the consequence of enhancing the measured asymmetry
($23.3 \pm 2.3 \%$ in KTeV~\cite{KTeVcoll}, $20 \pm 5 \%$ in NA48~\cite{NA48coll}).

4. Generalized Angular Distribution: As shown in Ref.~\cite{Heiliger:Sehgal},
a more complete study of the angular distribution of the decay
$K_L \rightarrow \pi^+ \pi^- e^+ e^-$ can yield further $CP$-violating observables,
some of which are sensitive to the non-radiative (charge-radius and short-distance)
parts of the matrix element. In particular the two-dimensional distribution
$d\Gamma / d \! \cos \theta_l d \phi$ has the form
\begin{eqnarray}
\frac{d\Gamma}{d\! \cos \theta_l d\phi} & = & K_1 + K_2 \cos 2 \theta_l +
K_3 \sin^2 \theta_l \cos 2 \phi + K_4 \sin 2 \theta_l \cos \phi \nonumber \\
 & & \mbox{} + K_5 \sin \theta_l \cos \phi + K_6 \cos \theta_l + K_7
\sin \theta_l \sin \phi \nonumber \\
 & & \mbox{} + K_8 \sin 2 \theta_l \sin \phi + K_9 \sin^2 \theta_l \sin 2 \phi.
\label{defKi}
\end{eqnarray}
Considering the behaviour of $\cos \theta_l$, $\sin \theta_l$, $\cos \phi$ and
$\sin \phi$ under $CP$ and $T$, the various terms appearing in Eq. 
(\ref{defKi}) have the following transformation:
\begin{center}
\begin{tabular}{|c|cc|}
\hline
& $CP$ & $T$ \\
\hline
$K_1$, $K_2$, $K_3$, $K_5$ & + & + \\
$K_4$, $K_6$ & $-$ & + \\
$K_8$ & + & $-$ \\
$K_7$, $K_9$ & $-$ & $-$ \\ \hline
\end{tabular}
\end{center}
Note that $K_{4,6,8}$ have $(CP)(T)=-$, a signal that they vanish in the
hermitian limit. If only the bremsstrahlung and magnetic dipole terms are
retained in the $K_L \rightarrow \pi^+ \pi^- e^+ e^-$ amplitude, one finds
$K_4 = K_5 = K_6 = K_7 = K_8 = 0$, the only non-zero coefficients being
$K_1 = 1$ (norm), $K_2 = 0.297$, $K_3 = 0.180$, $K_9 = -0.309$. In this
notation, the asymmetry in $d\Gamma / d\phi$ is $\mathcal{A} =
\frac{2}{\pi} \frac{\frac{2}{3} K_9}{1-\frac{1}{3}K_2} = - 14\%$. The
introduction of a charge-radius term induces a new $CP$-odd, $T$-even term
$K_4 \approx -1.3 \%$, while a short-distance interaction containing an axial
vector electron current can induce the $CP$-odd, $T$-odd term $K_7$. The
standard model prediction for $K_7$, however, is extremely small.\\

We conclude with a list of questions that could be addressed by future
research. In connection with $K_{L,S} \rightarrow \pi^+ \pi^- \gamma$: 
(i) Is there a departure from bremsstrahlung in 
$K_S \rightarrow \pi^+ \pi^- \gamma$
(evidence for direct $E1$)? (ii) Is there a $\pi^+ / \pi^-$ asymmetry in
$K_L \rightarrow \pi^+ \pi^- \gamma$ (evidence for $E2$)? (iii) Is there
a measurable difference between $\eta_{+-\gamma}$ and $\eta_{+-}$ (existence
of direct $CP$-violating $E1$ in $K_L \rightarrow \pi^+ \pi^- \gamma$)?
With respect to the decay $K_L \rightarrow \pi^+ \pi^- e^+ e^-$:
(i) Is there evidence of an $s$-wave amplitude? (ii) Is there evidence for
$K_4$ or $K_7$ types of $CP$-violation? On the theoretical front: (i) Can 
one calculate the $s$-wave amplitude, and
the form factors in $K_L \rightarrow \pi^+ \pi^- \gamma^*$~\cite{Savage}?
(ii) Can one understand the sign of $g_{M1}$? (iii) Can one explain why
direct $E1$ in $K_S \rightarrow \pi^+ \pi^- \gamma$ is so small compared
to direct $M1$ in 
$K_L \rightarrow \pi^+ \pi^- \gamma$ ($\left| g_{E1}/g_{M1} \right| < 5 \%$)?

\end{document}